\documentstyle[prl,aps,floats,epsf]{revtex}\def\narrowtext{}\tighten\twocolumn
\begin{document}
\draft

\title{
Zn-induced wipeout effect on Cu NQR spectra
in La$_{2-x}$Sr$_x$Cu$_{1-y}$Zn$_y$O$_4$
}
\author{
H. Yamagata,$^{1}$ H. Miyamoto,$^{1}$ K. Nakamura,$^{1}$ M. Matsumura,$^{1}$ and Y. Itoh,$^{2}$
}

\address{
$^1$Department of Material Science, Faculty of Science,\\ 
Kochi University, Kochi 780-8520, Japan\\
$^2$Japan Society for the Promotion of Science, Tokyo, Japan\\
}

\date{11 June, 2002}%
\maketitle %

\begin{abstract}%
We report a systematic study of Zn-substitution effect on Cu NQR spectrum
for high $T_c$ superconductors La$_{2-x}$Sr$_x$Cu$_{1-y}$Zn$_y$O$_4$ from
carrier-underdoped to -overdoped regimes (polycrystalline samples, $x$
=0.10, 0.15, and 0.20). We observed no appreciable wipeout effect for the
overdoped samples, a gradual and partial wipeout effect below about 80 K
for the optimally doped ones, and very abrupt and full wipeout effect
below about 40 K for the underdoped ones. The wipeout effect indicates a
highly enhanced spectral weight of Cu spin fluctuations at a low
frequency. We associate the wipeout effect with a Zn-induced local
magnetism far above 40 K and with a localization effect below 40 K. 
\end{abstract}

\narrowtext
  
Ultra slow spin fluctuation causes pair breaking effect and
suppresses the superconducting transition temperature $T_c$ of
high-$T_c$ cuprate superconductors \cite{Ohashi,Moriya}. The wipeout effect
on Cu nuclear quadrupole resonance (NQR) spectrum, which suggests the
existence of such a slow fluctuation, is observed in the deeply underdoped
La$_{2-x}$Sr$_x$CuO$_4$ \cite{Imai,Hunt} and in the nonmagnetic impurity
Zn-doped YBa$_2$Cu$_3$O$_7$ \cite{Yamagata0}. Thus, the wipeout effect on
Cu NQR spectrum may play a key role in understanding the mechanism of
suppression of $T_c$.     
 
Here, we report Zn-substitution effect 
on Cu NQR spectrum in La$_{2-x}$Sr$_x$Cu$_{1-y}$Zn$_y$O$_4$ from
carrier-underdoped to -overdoped regimes.
For Sr $x$=0.1, $T_c$ is $\sim$30, 13, $<$4.2 K for Zn $y$=0, 0.01, 0.02,
respectively.
For Sr $x$=0.15, $T_c$ is $\sim$38, 13 K for Zn $x$=0, 0.02, respectively. 
For Sr $x$=0.2, $T_c$ is $\sim$30, 13, $<$4.2 K for $y$=0, 0.03, 0.06,
respectively.   
We observed no appreciable wipeout effect for Sr-overdoped
samples, a partial wipeout effect below about 80 K for the optimally
doped ones, and full wipeout effect below about 40 K for
Sr-underdoped ones.  

 Zero field Cu NQR measurements with a pulsed spin-echo technique have
been carried out for powder samples, which were synthesized by a solid
state reaction method. The transverse relaxation curve of
the spin-echo was measured as a function of time $\tau$ between the
first and the refocusing pulses and was analyzed by a function of
$E(\tau)=E(0)$exp$[-2\tau/T_{2L}-0.5(2\tau/T_{2G})^2]$ with fitting
parameters $E(0)$, $T_{2L}$ and $T_{2G}$. The frequency integrated $E(0)$
is the observed intensity of NQR spectrum $I(x, y)$ (after $T_2$
correction). The details will be published elsewhere \cite{Yamagata1} 

Figure 1 shows the $T$ dependence of Cu NQR spectrum of Zn-free and
of Zn-doped samples with Sr-underdoped $x$=0.10 (left) and
-overdoped $x$=0.20 (right). Obviously, the NQR signal diminishes at
4.2 K for Sr-underdoped samples with Zn doping, whereas it is still
observable for Sr-overdoped ones with Zn doping. 

\begin{figure}
\epsfxsize=3.5in
\epsfbox{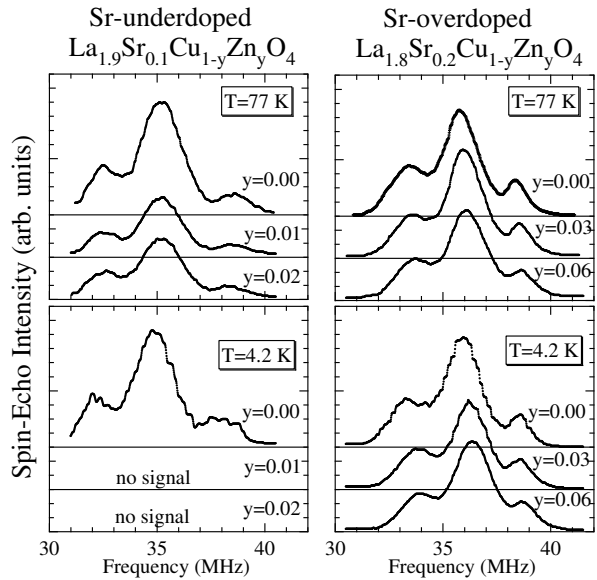}
\vspace{0.2cm}
\caption{
$T$ dependence of Cu NQR spectra in Zn-doped LSCO ($x$=0.1; $T_c\sim$30, 
13, $<$4.2 K for $y$=0, 0.01, 0.02) (left) and 
($x$=0.2; $T_c\sim$30, 13, $<$4.2 K for $y$=0, 0.03, 0.06) (right).}
\label{CuNQRZn}
\end{figure}

Figure 2(a) shows the $T$ dependence of the relaxation rate 1/$T_{2L}$($\gg
1/T_{2G}$ below about 100 K). No precursory effect, e.g.
divergence of 1/$T_{2L}$, is observed near the onset temperature of the
wipeout effect in the Zn-doped, Sr-underdoped sample ($x$, $y$)=(0.1,
0.02). Thus, the observed signal does not indicate any slowing down effect
of the observed spin fluctuation. A toy model of dynamical spin
susceptibility is shown in Fig. 2(b) to account for such a sudden wipeout
effect \cite{Yamagata1}.  

Figure 3 shows Zn-doping effect on the in-plane resistivity
(upper panel) \cite{Fukuzumi} and the relative intensity of the integrated
Cu NQR spectrum $I(x, y)/I(x, y=0)$ (lower panel). Since the resistivity is
metallic at 300 K, the high-$T$ wipeout effect, i.e.
$I(x, y)/I(x, y=0) < (1-y)$ at 300 K is associated with pure magnetic
effect due to the local field fluctuation of Zn-induced local moments
\cite{Xiao,Mahajan}. At lower temperatures the resistivity of
Sr-underdoped sample is semiconducting, so that the low-$T$ wipeout effect
is associated with electron localization effect. Thus, we associate
the wipeout effect with Zn-induced local magnetism far above 40 K and
with localization effect below 40 K.
  

\begin{figure}
\epsfxsize=3.5in
\epsfbox{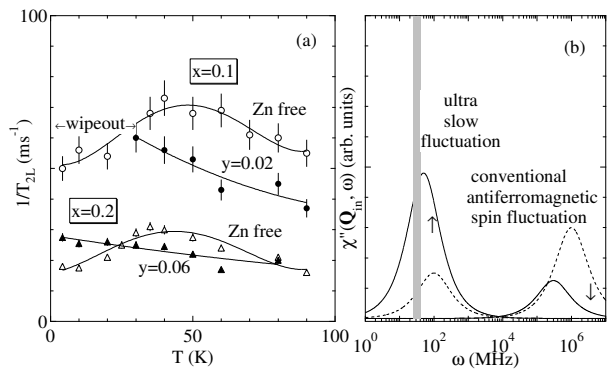}
\vspace{0.1cm}
\caption{
(a) $T$ dependence of the transverse relaxation rate 1/$T_{2L}$ for Zn-free
(open symbols) and Zn-doped samples (closed symbols). The solid curves are
guides for the eye. (b) A schematic toy model of dynamical spin
susceptibility to account for the wipeout effect without any precursory
divergence in 1/$T_{2L}$ nor in the Cu NQR linewidth. Ultra slow spin
fluctuation must appear to diminish suddenly the Cu NQR signal. 
The arrows indicate the direction of change when 
cooling down below the onset temperature of the wipeout
effect. The shaded area indicates an NQR frequency window around 35
MHz.     
}
\label{WipeoutT2}
\end{figure}

\begin{figure}
\epsfxsize=3.5in
\epsfbox{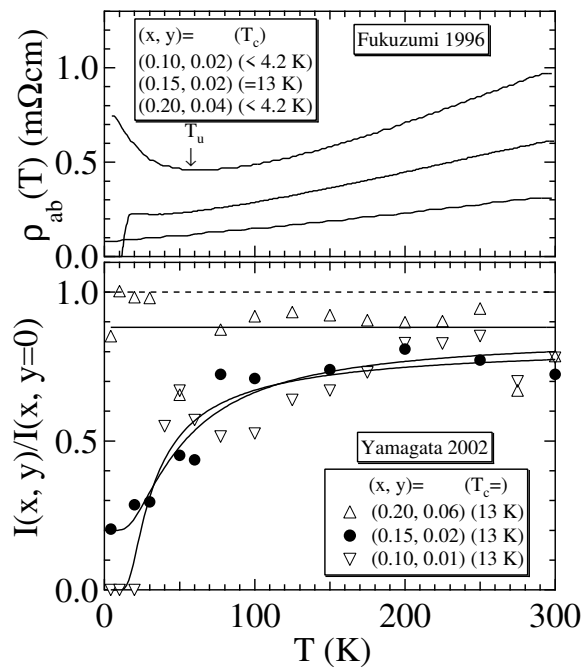}
\vspace{0.1cm}
\caption{
Zn-doping effect on the in-plane resistivity for single crystals,
reproduced from Ref. [7] (upper). 
The $T$ dependence of the ratio of the integrated intensity $I(x, y)$ of Cu
NQR spectrum of Zn-doped sample to Zn-free one (lower). Th solid curves
are guides for the eye. 
}
\label{WipeoutResistivity}
\end{figure}


\begin{references} 
 
\bibitem{Ohashi}Y. Ohashi, and H. Shiba, J. Phys. Soc. Jpn. {\bf 62}, 2783 (1993).

\bibitem{Moriya}T. Moriya, and K. Ueda, J. Phys. Soc. Jpn. {\bf 63}, 1871 (1994).

\bibitem{Imai}T. Imai, K. Yoshimura, T. Uemura, H. Yasuoka, and K. Kosuge, J. Phys. Soc. Jpn. {\bf 59}, 3846 (1990).

\bibitem{Hunt}A. W. Hunt, P. M. Singer, K. R. Thurber, T. Imai, Phy. Rev. Lett. {\bf 82}, 4300 (1999).

\bibitem{Yamagata0}H. Yamagata, K. Inada, and M. Matsumura, Physica C {\bf
185-189}, 1101 (1991).

\bibitem{Yamagata1}H. Yamagata, H. Miyamoto, K. Nakamura, M. Matsumura, and Y. Itoh, submitted to J. Phys. Soc. Jpn. (unpublished
work). 

\bibitem{Fukuzumi}Y. Fukuzumi, K. Mizuhashi, K. Takenaka, and S. Uchida, Phys Rev. Lett. {\bf 76}, 684 (1996).

\bibitem{Xiao}G. Xiao, M. Z. Cieplak, J. Q. Xiao, and C. L. Chien, Phys. Rev. B {\bf 42}, 8752 (1990).

\bibitem{Mahajan}A. V. Mahajan, H. Alloul, G. Collin, and J.-F. Marucco, Phys. Rev. Lett. {\bf 72}, 3100 (1994). 

\end{references}
\end{document}